\documentclass[pdflatex,sn-mathphys-num]{sn-jnl}

\usepackage{graphicx}%
\usepackage{multirow}%
\usepackage{amsmath,amssymb,amsfonts}%
\usepackage{amsthm}%
\usepackage[cal=rsfs]{mathalpha}
\usepackage[title]{appendix}%
\usepackage{xcolor}%
\usepackage{textcomp}%
\usepackage{manyfoot}%
\usepackage{booktabs}%
\usepackage{algorithm}%
\usepackage{algorithmicx}%
\usepackage{algpseudocode}%
\usepackage{listings}%
\usepackage{rotating}

\usepackage{aas_macros}

\theoremstyle{thmstyleone}%

\theoremstyle{thmstyletwo}%

\theoremstyle{thmstylethree}%

\begin{document}
\title[Characterization of near-Earth asteroid 2024~YR$_4$]{Rapid-response characterization of near-Earth asteroid 2024~YR$_4$ during a Torino Scale 3 alert}

\author*[1]{\fnm{Maxime} \sur{Devogèle}}\email{maxime.devogele@ext.esa.int} 

\author[2]{\fnm{Olivier R.} \sur{Hainaut}} 

\author[1]{\fnm{Marco} \sur{Micheli}} 

\author[3]{\fnm{Petr} \sur{Pravec}}

\author[4]{\fnm{Juan Luis} \sur{Cano}}

\author[1,5]{\fnm{Francisco} \sur{Ocaña}} 

\author[1]{\fnm{Luca} \sur{Conversi}}

\author[6]{\fnm{Nicholas} \sur{Moskovitz}} 

\author[7]{\fnm{Julia} \sur{de Le\'{o}n}}

\author[8]{\fnm{Zuri} \sur{Gray}} 

\author[8,9]{\fnm{Mikael} \sur{Granvik}} 

\author[18,8]{\fnm{Grigori} \sur{Fedorets}}

\author[10]{\fnm{Jules} \sur{Bourdelle de Micas}}

\author[10]{\fnm{Simone} \sur{Ieva}}

\author[10]{\fnm{Elisabetta} \sur{Dotto}}

\author[11]{\fnm{Tracie} \sur{Beuden}}

\author[11]{\fnm{Carson} \sur{Fuls}}

\author[6]{\fnm{Teddy} \sur{Kareta}}

\author[12]{\fnm{Stefano} \sur{Bagnulo}}

\author[13]{\fnm{Maria Antonella} \sur{Barucci}}

\author[14]{\fnm{Mirel} \sur{Birlan}}

\author[15]{\fnm{Andrea} \sur{Farina}}

\author[3]{\fnm{Kamil} \sur{Hornoch}}

\author[3]{\fnm{Petr} \sur{Fatka}}

\author[3]{\fnm{Peter} \sur{Ku\v{s}nir\'ak}}

\author[15]{\fnm{Francesca} \sur{Ferri}}

\author[13]{\fnm{Marcello} \sur{Fulchignoni}}

\author[15]{\fnm{Monica} \sur{Lazzarin}}

\author[15]{\fnm{Fiorangela} \sur{La Forgia}}

\author[10]{\fnm{Elena} \sur{Mazzotta Epifani}}

\author[15]{\fnm{Alessandra} \sur{Mura}}

\author[10]{\fnm{Davide} \sur{Perna}}

\author[16]{\fnm{Philippe} \sur{Bendjoya}}

\author[16]{\fnm{Jean-Pierre} \sur{Rivet}}

\author[17]{\fnm{Alberto} \sur{Cellino}}

\affil*[1]{%
  \orgdiv{ESA NEO Coordination Centre}, 
  \orgname{European Space Agency},
  \orgaddress{%
    \street{Largo Galileo Galilei, 1}, 
    \city{Frascati}, 
    \postcode{00044}, 
    \state{RM}, 
    \country{Italy}%
  }%
}

\affil[2]{%
  \orgdiv{European Southern Observatory}, 
  \orgname{ESO}, 
  \orgaddress{%
    \street{Karl-Schwarzschild-Straße 2}, 
    \city{Garching-bei-München}, 
    \postcode{85748}, 
    \state{Bavaria}, 
    \country{Germany}%
  }%
}

\affil[3]{%
  \orgdiv{Astronomical Institute of the Academy of Sciences of the Czech Republic}, 
  \orgaddress{%
    \street{Fri\v{c}ova 298}, 
    \city{Ond\v{r}ejov}, 
    \postcode{CZ-25165}, 
    \country{Czech Republic}%
  }%
}

\affil[4]{%
  \orgdiv{Planetary Defence Office}, 
  \orgname{ESA ESOC}, 
  \orgaddress{%
    \street{Robert-Bosch-Straße 5}, 
    \city{Darmstadt}, 
    \postcode{64293}, 
    \state{Hesse}, 
    \country{Germany}%
  }%
}

\affil[5]{%
  \orgdiv{ESA ESAC / PDO}, 
  \orgname{European Space Agency}, 
  \orgaddress{%
    \street{Bajo del Castillo s/n}, 
    \city{Villafranca del Castillo, Madrid}, 
    \postcode{28692}, 
    \country{Spain}%
  }%
}

\affil[6]{%
  \orgdiv{Lowell Observatory}, 
  \orgaddress{%
    \city{Flagstaff, AZ}, 
    \country{USA}%
  }%
}

\affil[7]{%
  \orgdiv{Instituto de Astrofísica de Canarias}, 
  \orgaddress{%
    \street{C/Vía Láctea s/n}, 
    \city{La Laguna},
    \postcode{38205}
    \country{Spain}%
  }%
}

\affil[8]{%
  \orgdiv{Department of Physics}, 
  \orgname{University of Helsinki}, 
  \orgaddress{%
    \street{P.O. Box 64}, 
    \postcode{00014}, 
    \city{Helsinki}, 
    \country{Finland}%
  }%
}

\affil[9]{%
  \orgdiv{Asteroid Engineering Laboratory}, 
  \orgname{Luleå University of Technology}, 
  \orgaddress{%
    \street{Box 848}, 
    \postcode{981 28}, 
    \city{Kiruna}, 
    \country{Sweden}%
  }%
}

\affil[10]{%
  \orgdiv{INAF}, 
  \orgname{Osservatorio Astronomico di Roma}, 
  \orgaddress{%
    \street{Via Frascati 33}, 
    \postcode{00040}, 
    \city{Monteporzio Catone (Roma)}, 
    \country{Italy}%
  }%
}

\affil[11]{\orgdiv{Lunar and Planetary Laboratory}, \orgname{University of Arizona}, \orgaddress{\street{1629 E. University Blvd.}, \city{Tucson}, \postcode{85721-0092}, \state{AZ}, \country{USA}}}

\affil[12]{%
\orgdiv{Armagh Observatory \& Planetarium}, 
  \orgname{}, 
  \orgaddress{%
    \street{College Hill}, 
    \postcode{BT61 9DB}, 
    \city{Armagh}, 
    \country{Northern Ireland, UK}%
  }%
}

\affil[13]{%
  \orgdiv{LIRA}, 
  \orgname{Observatoire de Paris, Université PSL, Sorbonne Université, Université de Paris-Cité, CY Cergy Paris Université, CNRS}, 
  \orgaddress{%
    \street{ 5 place Jules Janssen}, 
    \postcode{92195}, 
    \city{Meudon}, 
    \country{France}%
  }%
}

\affil[14]{%
  \orgdiv{IMCCE}, 
  \orgname{Observatoire de Paris, PSL Research University, CNRS UMR 8028, Sorbonne Université, Université de Lille}, 
  \orgaddress{%
    \street{77 av. Denfert-Rochereau}, 
    \postcode{75014}, 
    \city{Paris}, 
    \country{France}%
  }%
}

\affil[15]{%
  \orgdiv{Department of Physics and Astronomy ”Galileo Galilei”}, 
  \orgname{University of Padova}, 
  \orgaddress{%
    \street{Vicolo dell’Osservatorio 3}, 
    \postcode{35122}, 
    \city{Padova}, 
    \country{Italy}%
  }%
}

\affil[16]{%
  \orgdiv{Université Côte d’Azur}, 
  \orgname{CNRS}, 
  \orgaddress{%
    \street{OCA}, 
    \postcode{LAGRANGE}, 
    \city{Nice}, 
    \country{France}%
  }%
}

\affil[17]{%
  \orgdiv{INAF}, 
  \orgname{Osservatorio Astrofisico di Torino}, 
  \orgaddress{%
    \street{via Osservatorio 20}, 
    \postcode{101025}, 
    \city{Pino Torinese (TO)}, 
    \country{Italy}%
  }%
}

\affil[18]{%
  \orgdiv{Finnish Centre for Astronomy with ESO}, 
  \orgname{University of Turku}, 
  \orgaddress{%
    \street{Vesilinnantie }, 
    \postcode{20014}, 
    \city{Turku}, 
    \country{Finland}%
  }%
}


\abstract{\mathversion{normal}On 27 December 2024, near-Earth object (NEO) 2024\,YR$_4$ was discovered by the ATLAS survey and identified as a virtual impactor. A few weeks later, it eventually reached level 3 on the Torino Scale and was the first and only asteroid to be ever classified at that level. Here we report an intensive observational campaign combining time-series photometry in the visible, broadband visible and near-infrared colors, and low-resolution visible reflectance spectroscopy to assess its physical properties. Fourier analysis of the lightcurves yields a synodic rotation period of $P = 19.46341 \pm 0.00008$\,min, placing 2024\,YR$_4$ among the fast rotators, even if such rotation is common for objects of similar $H$ magnitude. Its visible and near-infrared colors and spectra are most consistent with an Sq or K taxonomic classification, though some ambiguity remains. Finally, its phase curve exhibits a notably shallow slope ($G = 0.51 \pm 0.11$), from which we derive an absolute magnitude of $H_\mathrm{R} = 23.82\pm0.09$\,mag. After color correction and taking into account other models for the phase function, we report an absolute magnitude of $H_\mathrm{V} = 24.14\pm0.25$\,mag. These characterizations, rotation period, taxonomy, and surface properties, would have been crucial for risk assessment and mitigation planning had the initially high impact probability scenario been confirmed, underscoring the importance for planetary defense of a rapid, coordinated international response.}

\keywords{2024 YR4, Near-Earth asteroids, Planetary defense, Photometry}



\maketitle

\section{Introduction}\label{sec:Introduction}
Planetary defense is an interdisciplinary field whose primary role is to protect Earth from the impact threat of comets and asteroids. It does not only involve astronomical observations and analysis but also complex political considerations like international decision-making, mitigation strategies, and disaster management planning. On the astronomical side, the main activities include discovery \citep[e.g.][]{Tonry_2018,Gregori_2023,Chambers_2016,Larson_1998}, follow-up \citep[e.g.][]{Vaduvescu_2013}, orbital analysis \citep[e.g.][]{Fenucci_2024,Chesley_2002}, and characterization \citep[e.g.][]{Thirouin_2018,Devogele_2019,Pravec_2014}.

NEOs are defined as any object in the Solar System (asteroids and comets) with a perihelion distance $q$ lower than 1.3 astronomical units (au). Throughout Earth's history, the impacts of NEOs have significantly shaped its evolution. Notably, 66 million years ago, a $\sim10$ km object impacted near the current Yucatán Peninsula resulting in the creation of the Chicxulub crater and the extinction of the dinosaurs \citep{Alvarez_1980,Schulte_2010}. More recently, in 1908, a $\sim 50$~m object impacted over the Tunguska region (Russia) and devastated over 2000 square kilometers of Siberian forest \citep{Napier_2009}. In 2013, the Chelyabinsk event resulted in the injuries of over 1500 people due to the entering in the atmosphere of an approximately 20\,m asteroid \citep{Borovivcka_2013}. 

Another significant event was the discovery of (99942)~Apophis in 2004. Apophis was initially assessed with a non-negligible impact probability for 2029 and reached Torino Scale\,4. The Torino Scale is a discrete and non-linear ranking from 0 to 10 that reflects both the probability of collision and potential damage from an impact. It is central to impact risk assessment and communication with the public \citep{Binzel_2000}. A Torino Scale of 0 means no risk (even in the case of an impact for very small objects) while 10 corresponds to an impact probability larger than 99\% for an object larger than 1\,km. Due to its Torino Scale 4, Apophis prompted extensive international observational and analytical efforts that eventually ruled out an Earth impact. However, even though it will not impact in 2029, it will have an exceptionally close fly-by to Earth as it will pass within about 38\,000 km, on 13 April 2029 \citep{Binzel_2009}. 

Impacts and close fly-bys are continuous reminders of the critical need for coordinated planetary defense strategies and international collaboration. Recognizing this, ESA created the Near-Earth Object Coordination Centre (NEOCC) in 2013 and its Planetary Defence Office (PDO) in 2019 while NASA established the Planetary Defense Coordination Office (PDCO) in 2016.

On 27 December 2024, asteroid 2024~YR$_4$ (hereafter YR4) was discovered by the Asteroid Terrestrial-impact Last Alert System (ATLAS) \citep{Tonry_2018} at the El Sauce Observatory, Chile. On the same day or shortly thereafter, the impact-monitoring systems, Aegis \citep{Fenucci_2024} at ESA/NEOCC, Sentry \citep{Chamberlin_2001} at NASA/JPL, and CLOMON2 \citep{Milani_2005} at NEODyS, all classified YR4 as a virtual impactor (i.e., an object with a non-zero impact probability in the next 100 years) and assigned it a Torino Scale of 1 for an impact on December 2032. The discovery of a new Torino Scale 1 object is not rare and currently occurs more or less once or twice per year. In most of the cases, these non-zero Torino Scale objects are short lived, and improvements of the orbit quickly remove all impact probabilities. However, this was not the case for YR4 as continued observations led to an increase of the impact probabilities, and by 27 January 2025, YR4 became the first object ever classified at Torino Scale 3 (Apophis was not and could never have been classified as 3 on the Torino Scale due to its larger size). This event triggered the first official notification by the International Asteroid Warning Network (IAWN), which issues alerts for objects larger than 10~m with impact probabilities exceeding 1\%. Finally, on 20 February 2025, as more observations were collected in the subsequent weeks, YR4 was first downgraded to Torino Scale 1, on 23 February 2025 to Torino Scale 0, and all the impact solutions for 2032 were removed on 8 March 2025.

The unique circumstances surrounding YR4, first ever Torino Scale 3 object, estimated size around 40 to 90 meters at the time (the JWST observations leading to an estimation of the size of $D=60 \pm 7$m \citep{Rivkin_2025} came after the observations presented in this work), and potential impact less than 10 years in the future required an urgent observational response.

In this paper, we present detailed photometric and spectroscopic observations obtained rapidly following its discovery, during the period when a potential impact scenario had not yet been ruled out and the impact probability was steadily rising. These observations were part of a global effort by the planetary defense community to gather as much information as they could to physically characterize the object. These physical properties include rotation period, surface composition, and brightness-phase behavior. These are instrumental to inform risk assessment and potential mitigation strategies in case the impact threat was confirmed.

YR4 will not be observable from ground-based observatories until a few months before its next close-approach to Earth in December 2028. If at that time we would have confirmed an impact with Earth in 2032, it would have most probably been too late to design deflection mission, like the DART mission \citep{Rivkin_2023}, or any other type of mitigation strategies to avoid the impact with Earth. Such mission needs to be planned and launched as early as possible and they would benefits from prior knowledge on the asteroid properties. 

\section{Observations}

We conducted a multi-instrument observational campaign to characterize the physical properties of YR4. These observations include time-series photometry for lightcurve analysis, visible and near-infrared broadband photometry for color determination, and visible spectroscopy for surface reflectance measurements. A summary of all observational epochs, organized by physical characterization method, is presented in Table~\ref{tab:obs_log}.

\begin{sidewaystable*}
\centering
\caption{Summary of physical properties observations of 2024~YR4. Observations are grouped by type: time-series (lightcurve), color photometry, phase curve, and spectroscopy. The same observations can be listed multiple times as the same dataset can be used for different characterization method.}
\label{tab:obs_log}
\begin{tabular}{llllllll}
\hline\hline
Date & Start-end & r & delta & $\alpha$ & Facility & Instrument & Filter (N. Exp.$\times$ Exp. Time)  \\
(UT) & (UT) & (au) & (au) & ($^\circ$) &  &  &  \\
\hline
\multicolumn{5}{l}{\textbf{Lightcurve observations}} \\
2024-12-28 & 10:53-11:01 & 1.005 & 0.026 & 35.0 & 1.0\,m CSS (I52) & 2k$\times$2k CCD & unfiltered (32$\times$8s)\\
2025-01-03 & 03:29–08:38 & 1.051 & 0.072 & 21.1 & 1.54\,m Danish & DFOSC & R (39$\times$75s)\\
2025-01-04 & 05:50–08:29 & 1.059 & 0.081 & 19.6 & 1.54\,m Danish & DFOSC & R (27$\times$75s)\\
2025-01-07 & 12:02-13:40 & 1.086 & 0.107 & 15.6 & 4.3\,m LDT        & LMI         & r (21$\times$30s)        \\
2025-01-21 & 02:43-03:23 & 1.207 & 0.225 & 7.4  & 8.2\,m VLT (UT1)     & FORS2       &  $R_{\rm SPECIAL}$ (45$\times$15s)    \\
\hline
\multicolumn{5}{l}{\textbf{Color photometry}} \\
2025-01-04 & 05:50–08:29 & 1.059 & 0.081 & 19.6 & 1.54\,m Danish  & DFOSC & R (27$\times$75s), V (6$\times$75s)\\
2025-01-07 & 12:02-13:40 & 1.086 & 0.107 & 15.6 & 4.3\,m LDT           & LMI         & g (6$\times$30s), $r$ (21$\times$30s),\\
 & &  &  &  &  &        &  $i$ (6$\times$30s), $z$ (12$\times$30s) \\
2025-01-10/11 & 23:33 - 01:13 & 1.116 & 0.136 & 12.1 &3.6\,m TNG           & DOLORES     & B (12$\times$90s), V (12$\times$90s), \\
& & & & & & & R (12$\times$90s), I (12$\times$90s)\\
2025-01-21 & 04:15-06:03  & 1.208 & 0.226 & 7.4 & 8.2\,m VLT (UT4)     & HAWK-I      & Y (30$\times$60s), $J$ (16$\times$60s),  \\
 & &  &  &  &  &        & $H$ (26 $\times$ 60s)  \\
\hline
\multicolumn{5}{l}{\textbf{Phase curve photometry}} \\
2025-01-03 & 03:29–08:38 & 1.051 & 0.072 & 21.1 & 1.54\,m Danish & DFOSC & R (39$\times$75s)\\
2025-01-04 & 05:50–08:29 & 1.059 & 0.081 & 19.6 & 1.54\,m Danish & DFOSC & R (27$\times$75s)\\
2025-01-21 & 02:43-03:23 & 1.207 & 0.225 & 7.4  & 8.2\,m VLT (UT1)     & FORS2       & $R_{\rm SPECIAL}$ (45$\times$15s)     \\
2025-01-31 & 00:46-01:40 & 1.301 & 0.323 & 11.2 & 2.5\,m NOT           & ALFOSC      & R (26$\times$120s)              \\
2025-02-01/02 & 23:30-00:20 & 1.319 & 0.344 & 12.2 & 2.5\,m NOT      & ALFOSC      & R (13$\times$240s)             \\
2025-02-05 & 00:36-01:27 & 1.348 & 0.377 & 13.8 & 2.5\,m NOT           & ALFOSC      & R (11$\times$300s)              \\

\hline
\multicolumn{5}{l}{\textbf{Spectroscopy}} \\
2025-01-13 & 01:50-02:31 & 1.134 & 0.154 & 10.4 & 10.4\,m GTC          & OSIRIS+      & R500R, 0.48-0.92 $\mu$m (3$\times$1200s)  \\

\hline
\end{tabular}
\end{sidewaystable*}
Time–series photometry of YR4 began with two nights of R-band and one night of V-band exposures obtained at the 1.54\,m Danish Telescope (DK154; W74) on 3 and 4 January 2025. The observations were carried out by Petr Pravec and his colleagues within their NEA light-curve photometry programme NEOSource\footnote{\url{https://space.asu.cas.cz/\textasciitilde ppravec/}} and supported by the Czech Academy of Sciences through its {\it Praemium Academiae} award. Information on the instrument and telescope characteristics can be found in \citet{Fatka_2025}. Observing procedures and data reduction follow those described in \citet{Pravec_2014}.

Photometric observations were then obtained at the 4.3\,m Lowell Discovery Telescope (LDT; G37), Happy Jack, Arizona through time allocated to PI Nicholas Moskovitz as part of the Mission Accessible Near-Earth Object Survey (MANOS) program \citep{Moskovitz_2014,Thirouin_2016,Thirouin_2018,Devogele_2019}. These observations were performed in the SDSS g'-, r'-, i'-, and z'-bands with the Large Monolithic Imager (LMI) \citep{Massey_2013}. We used similar observation strategies as those used in \citet{Devogele_2021}. These observations were used for both color measurements and lightcurve analysis (see Sect.~\ref{sec:lightcurve}).

Additional visible photometry was acquired through Director's Discretionary Time (DDT) at the 3.58\,m Telescopio Nazionale Galileo (TNG; Z19) at the Roque de los Muchachos Observatory (La Palma, Spain), awarded under a pilot program (AOT50\_31) in support of the European Commission-funded NEOPOPS project. Observations were obtained with the DOLORES instrument using Johnson–Cousins B-, V-, R-, and I-filters. 
Flat-field files were taken as close as possible in time to the observing nights, while bias frames were taken prior to the observations on the same night.

Observations in both the visible and near-infrared were obtained using the 8.2\,m Very Large Telescope (VLT; 309) at Cerro Paranal, Chile. The R-band photometry was obtained with the FORS2 instrument \citep{Appenzeller_1998} on Unit Telescope 1 (UT1, Antu) as part of a long-standing ESA–ESO partnership program dedicated to the follow-up of potential impactors for planetary defense (PI: Olivier Hainaut). Almost simultaneously, near-infrared observations were acquired with the HAWK-I instrument \citep{KisslerPatig_2008} on Unit Telescope 4 (UT4, Yepun) through a Director’s Discretionary Time program (PI: Maxime Devog\`{e}le), submitted on 19 January, awarded on 20 January, and executed during the night of 20-21 January 2025.

The FORS2 observations were obtained using a binning factor of 2$\times$2 and with the $R_{\rm SPECIAL}$ filter and reduced using the ESO FORS2 pipeline \citep{Freudling_2013}. Photometric calibration to the Cousins R band was achieved using SDSS DR9 reference stars. In particular, the SDSS DR9 catalog's g, r, i, z magnitudes were converted to R-band magnitudes using the empirical color transformations of \citet{Chonis_2008}, which apply first‐order color terms to account for filter differences and deliver R‐band zero points accurate to $\sim$0.02 mag. As a sanity check, we also calibrated the FORS2 data using the Pan-STARRS catalog that is also defining a relation to transform from the SLOAN bands to the R band \citep{Tonry_2012}. The median differences between all the observations calibrated with the both catalogs is only $\sim 0.01$ magnitude.

HAWK-I imaging was performed in the Y-, J-, and H-broadband filters. The FORS2 R-band data were used to connect the visible and NIR color measurements.

Low-resolution visible spectroscopy was obtained with the OSIRIS+ instrument \citep{2000SPIE.4008..623C,2010ASSP...14...15C} on the 10.4\,m Gran Telescopio Canarias (GTC), located at the Roque de Los Muchachos Observatory, under program GTC16-24B (PI: Julia de León). This is a regular allocation for the characterization of NEOs accessible to spacecraft \citep[NHATS as defined by][]{Abell_2012}, potentially hazardous asteroids or PHAs (i.e., objects with Minimal Orbital Intersection Distance, MOID, lower than 0.05 au and an absolute magnitude $H$ larger than 22), and virtual impactors. The R500R grism and a $1.2''$ slit width were used, yielding a spectral resolution of $R=587$ for a $0.6''$ slit and a dispersion of 4.88\,\AA/pixel, covering the 0.48–0.92\,$\mu$m wavelength range. Observations were aligned along the parallactic angle and tracked at the asteroid's apparent motion. The object was at apparent magnitude of $V$=20.8 at the moment of the observations so three 1200\,s individual exposures were obtained with a 10'' offset along the slit. The solar analog stars SA98-978 and SA102-1081 were also observed to correct for the solar spectrum. 

Additional photometric measurements were obtained with the ALFOSC (Andalucia
Faint Object Spectrograph and Camera) instrument on the 2.5\,m Nordic Optical Telescope (NOT) using a Bessel R filter. ALFOSC is equipped with a $2048\times2048$ CCD providing a $6.4'\times6.4'$ field and $0.2138''$/pixel scale. These observations were conducted as target of opportunity (ToO) observations under the long-running ``ESA S2P Program'' (proposal ID 68-803), dedicated to monitoring NEOs in the context of planetary defense.

Finally, archival time-series photometry from the Catalina Sky Survey (CSS) \citep{Christensen_2012} were retrieved. These observation had been obtained on 28 December 2024 with the 1.0\,m telescope located on Mount Lemmon (MPC code I52).

\section{Lightcurve}\label{sec:lightcurve}

We analyzed time-series photometric data of YR4 obtained with four different facilities: the 1\,m I52 telescope from the Catalina Sky Survey (CSS), the 1.5\,m Danish telescope (DK154), the 4.3\,m Lowell Discovery Telescope (LDT), and the 8.2\,m Very Large Telescope (VLT). We did not consider the Nordic Optical Telescope observations, as the signal was too low to extract time-resolved photometry; usable photometry could only be obtained after stacking images spanning more than one rotation period. The Telescopio Nazionale Galileo observations were acquired solely for color measurements and, being taken in different filters with too few exposures per filter, did not allow the extraction of a lightcurve. A summary of the observing conditions and setups is given in Table~\ref{tab:obs_log}.

\subsection{Danish 1.54 m telescope observations}

\begin{figure}[t!]
\centering
\includegraphics[width=14cm]{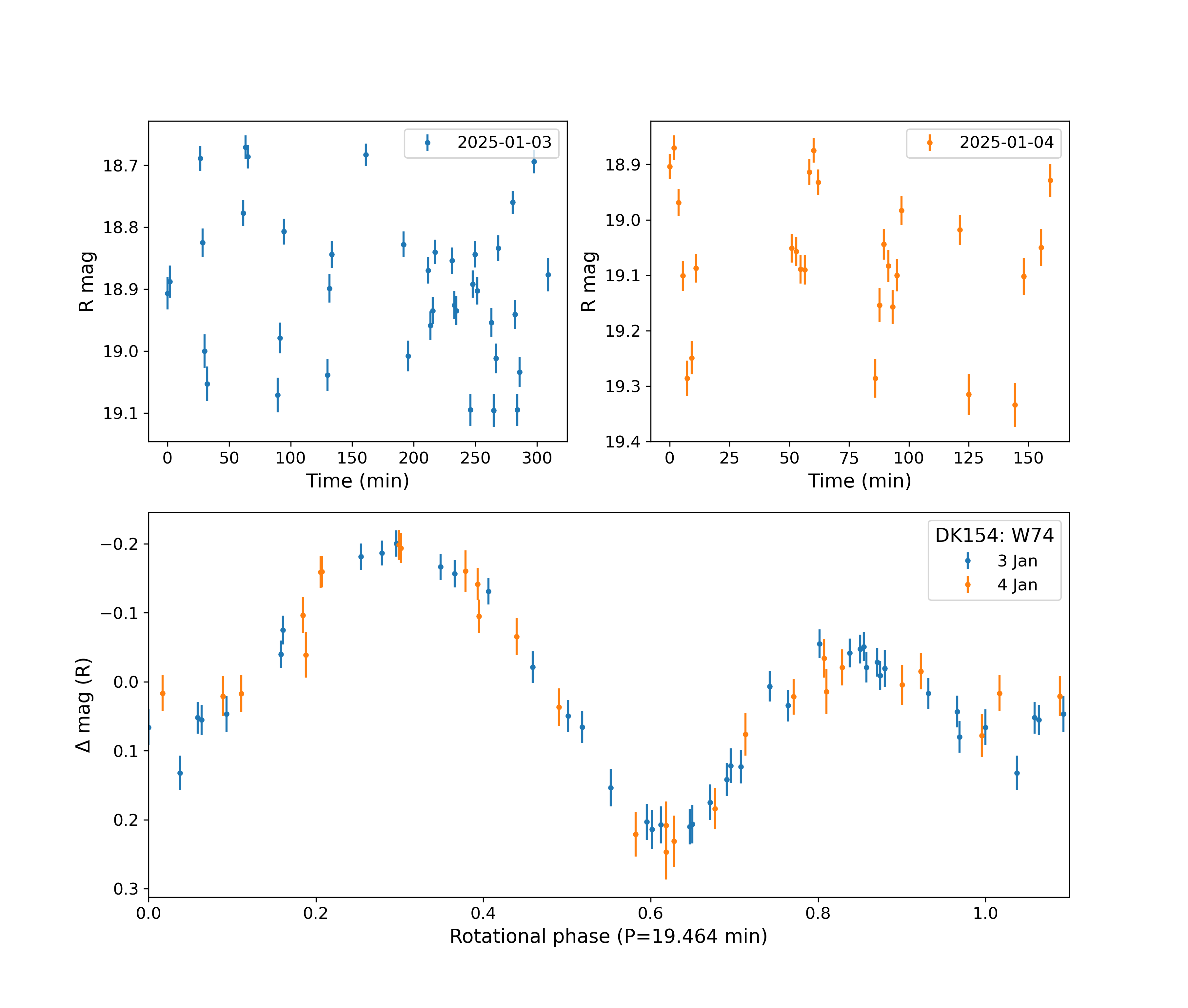}
\caption{Top row: Lightcurves obtained at the Danish 1.54\,m (DK154; W74) Telescope on 3 and 4 January 2025. Bottom panel: Phased lightcurve of the two DK154 datasets according to a rotation period of 19.464 minutes.}
\label{fig:DK154_LC}
\end{figure}

Time-series photometry was obtained with the DFOSC camera on the 1.54\,m Danish Telescope (DK154) at ESO La Silla (Chile). Two R-band observations were obtained on 3 and 4 January 2025, yielding 39 and 27 individual exposures, respectively, each of 75 seconds. The sequences cover $\sim\!5.1$ hours on the first night and $\sim\!2.6$\ hours on the second (Fig.~\ref{fig:DK154_LC}). The standard Cousins R-band filter were used to conduct the observations. The absolute calibrations were done with Landolt standard stars \citep{Landolt_1992}. The telescope was tracked at half of the apparent rate of the asteroid, providing star and asteroid images of the same profile in one frame. The data were reduced using standard procedures (bias subtraction and flat-fielding) and photometrically reduced with the aperture photometry reduction software package {\it Aphot32}. Analysis of the two datasets from the DK154 telescope gives an unambiguous rotation period of $P=19.464 \pm 0.002$ minutes for YR4.

\subsection{LDT 4.3 m observations}

\begin{figure}[t!]
\centering
\includegraphics[width=8cm]{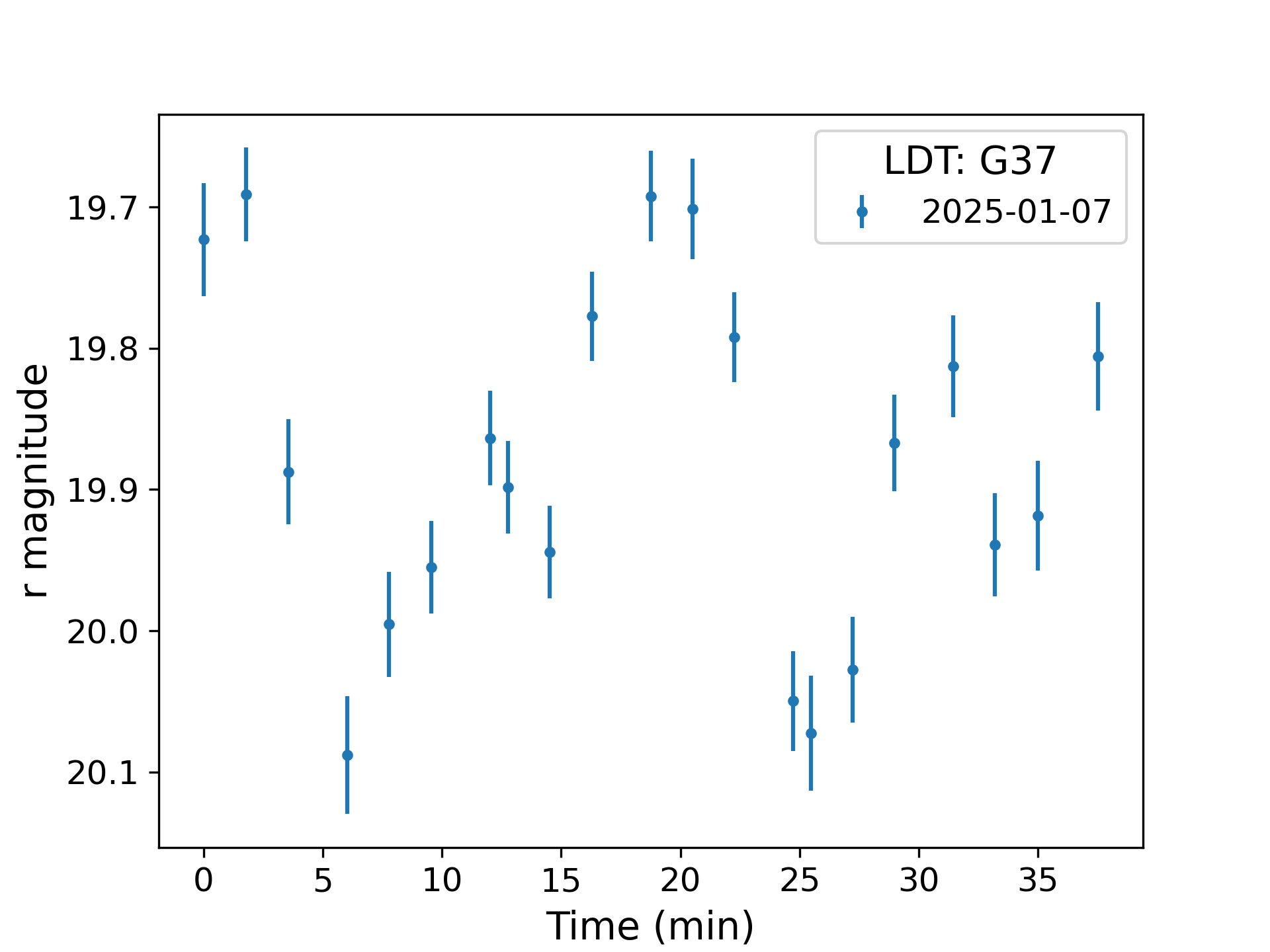}
\caption{Lightcurve obtained at the 4.3\,m Lowell Discovery Telescope (LDT; G37). 21 individual acquisitions of 30 seconds each were obtained spanning a total time of 37 minutes.}
\label{fig:LDT_LC}
\end{figure}

The LDT r-band dataset consists of 21 exposures of 30 seconds each, obtained over a total span of 37 minutes on 7 January 2025. The data were reduced using standard procedures (bias subtraction and flat-fielding), and photometry was extracted using the \texttt{photometrypipeline} \citep{Mommert_2017}. Calibration was performed against the Pan-STARRS catalog. The resulting lightcurve displays two well-defined minima and maxima, suggesting a rotation period near 37 minutes (see Fig.~\ref{fig:LDT_LC}). However, from the DK154 datasets we know that the alternating (second) minimum and maximum are intrinsically shallower; in the LDT sequence they are not discernible because of the lower time resolution of the data compared to the DK154 dataset. 

\subsection{VLT observations}

\begin{figure}[t!]
\centering
\includegraphics[width=8cm]{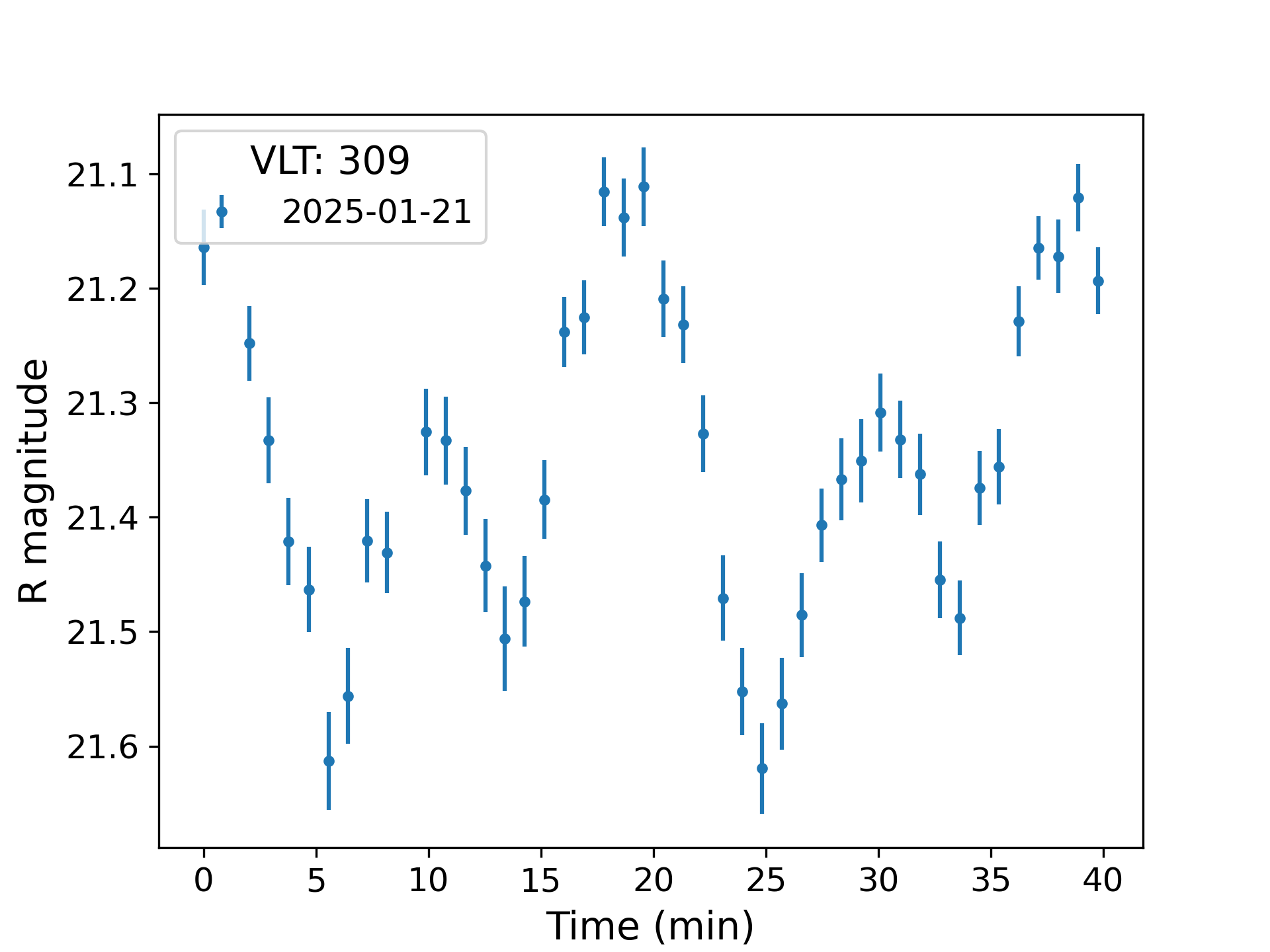}
\caption{Lightcurve obtained at the 8.2\,m Very Large Telescope (VLT; 309). 45 individual acquisitions of 15\,s each were obtained spanning a total time of 40 minutes}
\label{fig:VLT_LC}
\end{figure}

We observed YR4 on 21 January 2025 using the FORS2 instrument on the 8.2\,m VLT (UT1/Antu). The dataset consists of 45 individual exposures of 15 seconds, acquired over 40 minutes (Fig.~\ref{fig:VLT_LC}). Despite the target being $\sim$1.5\,mag fainter than during the LDT observations, the larger aperture of the VLT allowed shorter exposure times and higher signal-to-noise ratio (SNR). Data were reduced using the ESO FORS2 pipeline and calibrated in the R band with the SDSS DR9 catalog. The higher cadence and quality of this dataset enables a more precise resolution of the asteroid's lightcurve, revealing the periodicity near 19.5 minutes (see Fig.~\ref{fig:VLT_LC}) already observed in the DK154 datasets.

\subsection{Catalina Sky Survey data}

\begin{figure}[t!]
\centering
\includegraphics[width=8cm]{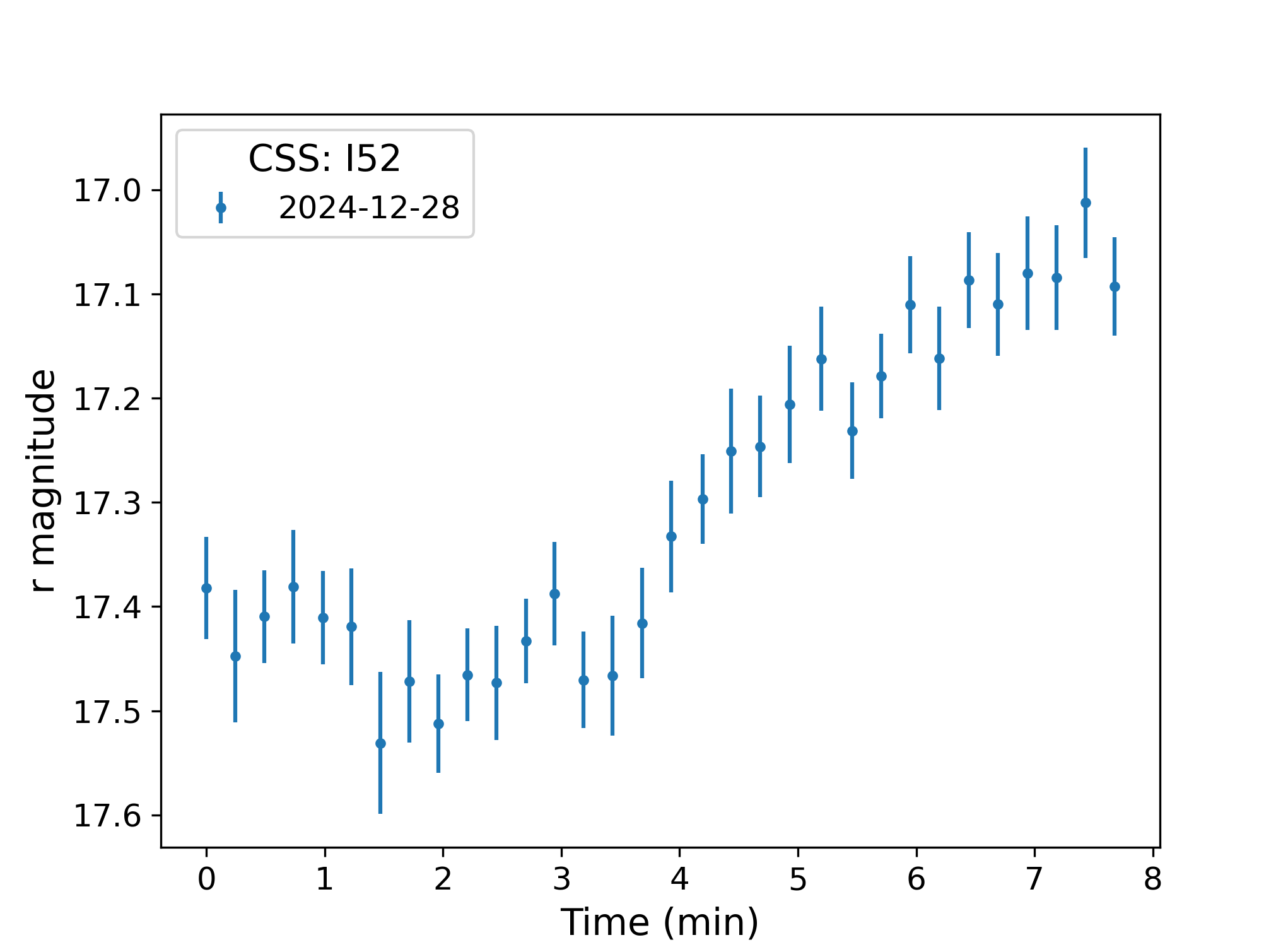}
\caption{Lightcurve obtained at the 1\,m Mount Lemmon telescope (I52). 32 individual acquisitions of 8\,s each were obtained spanning a total time of 8 minutes}
\label{fig:I52_LC}
\end{figure}

To supplement our observations, we retrieved archival data from the Catalina Sky Survey (CSS) obtained on 2024 December 28 using the 1\,m I52 telescope at Mount Lemmon. This dataset consists of 32 unfiltered exposures of 8 seconds each, spanning a total duration of approximately 8 minutes (Fig.~\ref{fig:I52_LC}). Although the observation window is shorter than the asteroid's rotation period, the high cadence and signal-to-noise ratio proved valuable for phasing with other datasets. We performed the reduction and photometric calibration ourselves using the SDSS DR9 catalog to transform the unfiltered magnitudes to the r band.

\subsection{Rotation period analysis}
\label{sec:rotation_anal}
To determine the synodic rotation period of YR4, we used a Fourier series fitting approach applied to time-series photometry obtained at four different epochs. The lightcurve model consists of a truncated Fourier series with $N$ harmonics, of the form:

\begin{equation}
f(t) = a_0 + \sum_{n=1}^{N} \left[ a_n \cos\left(\frac{2\pi n t}{P}\right) + b_n \sin\left(\frac{2\pi n t}{P}\right) \right],
\end{equation}

where $P$ is the fixed trial rotation period. For each tested period, the coefficients $a_n$ and $b_n$ were optimized using the Levenberg–Marquardt algorithm via the \texttt{scipy.optimize.curve\_fit} routine \citep{2020SciPy}. The best-fit period was identified by minimizing the reduced $\chi^2$ across all datasets. 

To identify acceptable period solutions, we adopted the standard $\Delta\chi^2$ approach for confidence interval estimation, retaining only those candidate periods with $\chi^2$ values within the 3$\,\sigma$ threshold of the best-fit solution. Specifically, for one estimated parameter (since the rotation period is fixed at each step and only the Fourier series coefficients are optimized, only the rotation period is considered as an ``interesting parameter'' while the Fourier coefficients are considered as ``uninteresting parameters'' \citep[as defined by][]{Avni1976}, this corresponds to $\Delta\chi^2 = 9.0$. For the acceptable solutions, the quoted uncertainties are the 1$\,\sigma$ uncertainties corresponding to $\Delta\chi^2 = 1.0$

The initial search was conducted with the two R-band light curves obtained at the DK154 telescope on 3 and 4 January 2025. The first sequence comprises 39 exposures acquired over 5.1 hours (03:29–08:38 UT), and the second 27 exposures spanning 2.6 hours (05:50–08:29\,UT) on the following night. A Fourier fit to the combined DK154 yielded a single, well-defined synodic period of $P = 19.464 \pm 0.002$\,min, with no statistically significant secondary minima in the periodogram.

To test the uniqueness and improve on this solution, we added the LDT sequence of 7 January 2025 and the VLT/FORS2 sequence of 21 January 2025, obtained 4 days and 18 days after the Danish observations, respectively. Although these additional data were separated from the DK154 epoch by $\sim300$ and $\sim1100$ rotations, folding them on the DK154 period produced an excellent phase match with no measurable drift.  A joint Fourier fit across the three datasets refined the period to $19.4634 \pm 0.0001$\,min.

Finally, we incorporated the archival I52 CSS light curve from 28 December 2024. Despite its short 8 minutes span and thus its partial observation of YR4 lightcurve, we were able to link the phase between the December and January observations and significantly improve the period solution. The global minimum in the profile-$\chi^{2}$ curve occurs at $P = 19.46341 \pm 0.00008$\,min. All alternate aliases lie beyond the $3\sigma$ $(\Delta\chi^{2}=9)$ threshold for a single parameter.

\begin{figure}[t!]
\centering
\includegraphics[width=8cm]{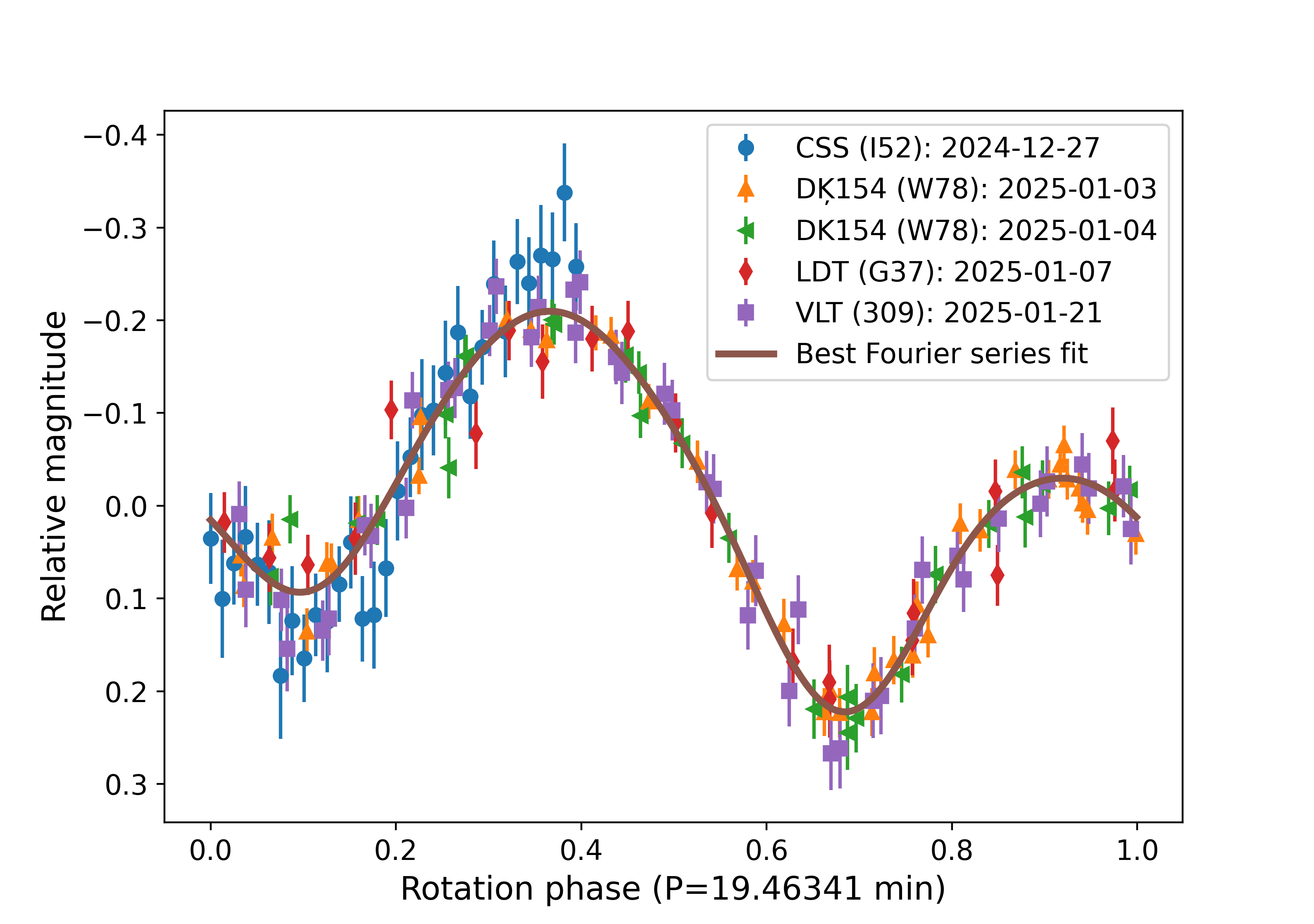}
\caption{Phased lightcurve of YR4 according to a period of 19.46341 minutes. The best Fourier series fit of order 5 is plotted as a continuous line.}
\label{fig:Lightcurve}
\end{figure}

The phased composite light curve (Fig.~\ref{fig:Lightcurve}) shows a peak-to-trough amplitude of 0.43\,mag. Taking into account that our VLT observations were obtained at a phase angle of $7.4^{\circ}$ and \citet{Zappala_1990} amplitude correction to opposition (we are here assuming a typical phase-amplitude correction of 0.03 mag/$^{\circ}$), we find an expected light-curve amplitude at opposition of $\sim0.36$\,mag. Interpreted in terms of a triaxial ellipsoid, this amplitude implies an upper limit on the equatorial axis ratio of $b/a \sim 0.72$ \citep{Surdej_1978}.

\section{Phase curve}

To characterize the absolute magnitude $H$ of YR4, we constructed a phase curve using photometry obtained exclusively in the R band to avoid complications from color transformations and phase slope variability across filters. This dataset includes observations from DFOSC on the DK154, FORS2 on the VLT, and ALFOSC on the NOT.

\subsection{Photometric methodology}

Because YR4's light curve is displaying an amplitude of 0.43\,mag, the rotational signal must first be removed before individual nights can be combined into a reliable phase curve. We perform this removal using the Fourier series model, obtained using all the available light curves, that was derived in Sect.~\ref{sec:rotation_anal} and displayed in Fig.\,\ref{fig:Lightcurve}. For each set of exposures used for the phase-curve modeling and for which we have obtained a light curve (DK154 and VLT), we evaluated the best-fit Fourier series for each exposures and subtracted the modeled magnitude from the measured one. Since the phase angle does not vary significantly over the timespan of individual nights (at most 0.002$^{\circ}$ for both DK154 datasets), we average all observations within each set into a single photometric measurement. This approach prevents unnecessary scatter at identical phase angles due to residual light-curve correction errors and avoids introducing biases in the fit arising from unequal number of measurements at different phase angles. The quoted nightly magnitude is therefore the mean model-corrected value, and its uncertainty is the standard deviation of the residuals within that light curve. Similar approaches have been used by several authors \citep[e.g.][]{Waszczak_2015,Kwiatkowski_2021,Terai_2013} to correct for the lightcurve-induced scatter in the phase-curve analyses. \citet{Waszczak_2015} analyzed a large number of asteroids and showed that modeling and removing the rotational variation with a Fourier series reduces the photometric root-mean-square scatter by a factor of about 3.

The NOT observations were obtained over three nights and consisted of low SNR individual exposures and light curves could not be derived from them. To improve the photometric precision, we applied a two-step stacking strategy: first, sidereal stacks were used to calibrate the zero point using field stars; second, the same images were re-stacked at the asteroid's non-sidereal rate to extract the asteroid flux. Since both stacks are derived from the same exposures, we assume a common photometric zero-point. This method effectively averages out rotational brightness variations over the one-hour sequences. Data from the final NOT night (4 February 2025) were discarded due to contamination by a nearby field star.

\subsection{Phase curve fitting}

We fit the resulting photometry using the $H,G$ photometric phase function from \citet{Bowell1989}, the $H,G_{12}^{*}$ formulation from \citet{Penttila_2016}, and the $H,G_{1},G_{2}$ model from \citet{Muinonen_2010} based on the implementations provided by the {\tt sbpy} photometry package \citep{Mommert_2019}. For the first two  models, we evaluated the fit over a two-dimensional grid of parameter values, with the slope parameter ranging from 0 to 1 and the absolute magnitude $H_\mathrm{R}$ from 22 to 26. At each grid point, we computed the model prediction and corresponding $\chi^2$ value. The best-fit solution was identified as the grid point minimizing the $\chi^2$, and the 1\,$\sigma$ confidence region in the $H$-slope parameter space was determined using the $\Delta\chi^2 = 2.3$ criterion appropriate for two jointly fitted parameters \citep{Avni1976}.

\begin{figure}[t!]
\centering
\includegraphics[width=8cm]{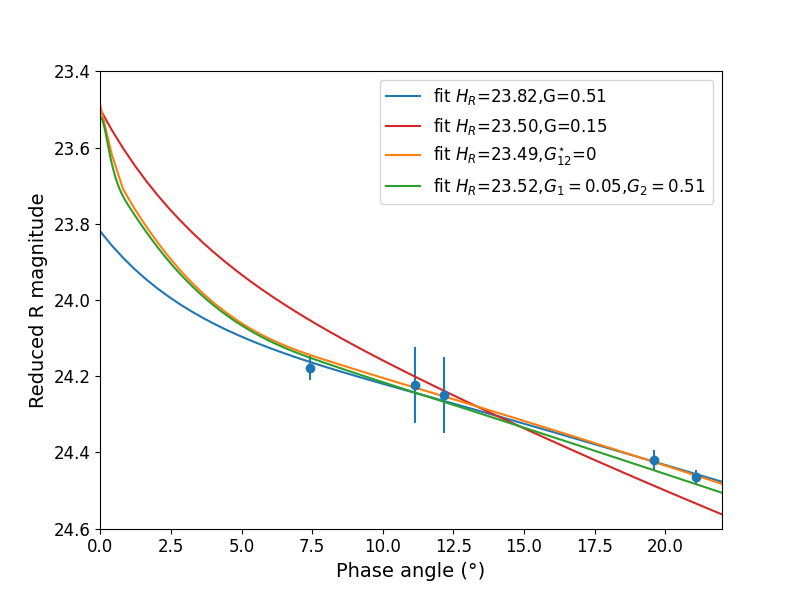}
\caption{Phase curve of YR4 in the R band. In blue is shown the best fit with a phase slope of $G = 0.51$ and an $H_{\rm R}$ = 23.82\,mag. In red is shown the best fit with a phase slope of forced $G = 0.15$. In orange is shown the best fit using the $H,G_{12}^{*}$ model with $H_{\mathrm R} = 23.49$\,mag and $G_{12}^{*} = 0$. In green is shown the best fit using the $H,G_1,G2$ model with $H_{\mathrm R} = 23.52$\,mag, $G_{1}= 0.05$, and $G_{1}= 0.52$. }
\label{fig:Phase_Curve}
\end{figure}

For the  $H$, $G$ model the best-fit solution yields $G = 0.51 \pm 0.11$ and $H_\mathrm{R} = 23.82 \pm 0.09$\,mag. This corresponds to a very shallow phase slope compared to the typical $G = 0.15$ often assumed for asteroids. A fit forced to $G = 0.15$ results in a significantly poorer match to the data and $H_\mathrm{R} = 23.5$\,mag, as shown in Fig.~\ref{fig:Phase_Curve}.

Using the $H,G_{12}^{*}$ phase function results into a similarly good fit to the data, but a significantly different $H_\mathrm{R}$ value of $23.49 \pm 0.02$\,mag. This is due to the fact that we are lacking observations at low phase angles and we cannot constrain the effect of the opposition surge. 

Fitting the single-parameter $H,G_{12}^{\ast}$ phase law to our data drives the solution to $G_{12}^{\ast}=0$; the best solution actually yields a negative value, which we clip to the allowed lower bound. \citet{Mahlke_2021} showed that such boundary solutions $(G_{12}^{\ast}=0$ or 1) are common when phase-angle coverage is limited, because the single parameter cannot simultaneously reproduce both the absolute magnitude and the slope of the curve. This limitation of the model is also reflected in the under-estimated uncertainty on the $H_\mathrm{R}$ magnitude. Our observations start at $\alpha\simeq7^{\circ}$ and never sample the opposition surge, so the low-phase behavior remains unconstrained. 

For the $H,G_{1},G_{2}$, we used an MCMC approach with prior information on the $G_1$ and $G_2$ values based on previous fit of hundreds of thousands of asteroid phase function of asteroids \citep{Mahlke_2021}, and following the procedure outlined in \citet{Vega_2023}. The best fit provides $H_\mathrm{R}=23.52 \pm 0.10\,$mag, $G_1 = 0.05 \pm 0.04$, and $G_2 = 0.51 \pm 0.04$ which is consistent with the results obtained with the $H,G_{12}^{\ast}$ model.

We find that the $HG$ phase-function yield incompatible solutions for the absolute magnitude $H$ compared to the $H,G_{12}^{\ast}$ and \(H,G_{1},G_{2}\) models. As a result, the derived value of $H_{\rm R}$ should be considered preliminary; additional observations at low phase angles are required to constrain a physically meaningful phase-function fit. In the meantime, we adopt a conservative estimate of $H_{\mathrm R} = 23.66 \pm 0.25\,$mag, corresponding to the average of the best-fit solutions from all models, with an uncertainty encompassing the full range of their respective $1\,\sigma$ confidence intervals.

From our DK154 V‐ and R-band observations, we derive a color index of $(V - R) = 0.48 \pm 0.02\,$mag (see Sect. \ref{sec:tax}) in the Johnson-Cousins photometric system. Combined with the R‐band absolute magnitude estimate of $H_{\mathrm{R}} = 23.66 \pm 0.25\,$mag, this color corresponds to a V‐band absolute magnitude of $H = 24.14 \pm 0.25\,$mag.

\section{Taxonomy}
\label{sec:tax}
To constrain the taxonomic type of YR4, we used a combination of visible and near-infrared color photometry as well as low-resolution visible spectroscopy. We produced a composite reflectance spectrum and evaluated its compatibility with standard taxonomic classes.

\subsection{Photometric reflectance spectrum}

At the DK154, observations were obtained in both R and V-band. As presented in Sect. \ref{sec:lightcurve}, the R-band observations were used to construct a light curve of YR4 over two consecutive nights. During the night of 4 January 2025, V-band observations were interleaved in between R-band measurements. This allows for an accurate determination of the Johnson-Cousins (V-R) color index by taking into account the light-curve variation. The DK154 observations resulted in a ($\mathrm{V}-\mathrm{R}) = 0.48 \pm 0.02$\,mag

The LDT observations were conducted in the SDSS g', r', i', and z' filters using the LMI instrument. Photometry was extracted from bias- and flat-field-corrected images, calibrated using the Pan-STARRS catalog \citep{Magnier_2020}, and converted to reflectance by normalizing to solar absolute magnitude. We are here using the following absolute magnitude for the Sun $M_{\odot,g}=5.12$, $M_{\odot,r}=4.68$, $M_{\odot,i}=4.57$, $M_{\odot,z}=4.54$. These values have been derived based on the eighth SDSS-III data release \citep{Aihara_2011}, filter transformations from \citet{Bilir_2005}, and colors transformation from \citet{Rodgers_2006}. The observations consisted of three consecutive sequences of r'g'r'i'r'z'z' exposures. The observations obtained in the r' filter were used to model the lightcurve variability of YR4, providing a reference to correct the flux variations in the other filters. The z' filter was repeated to increase the signal-to-noise ratio, as observations in that band provide lower fluxes.

The DOLORES observations at the TNG were performed in the Johnson–Cousins B, V, R, and I filters. Standard aperture photometry was applied, and calibration was done using field stars with known magnitudes from the SDSS DR9 catalog, following standard transformations and employing the \texttt{photometrypipeline} \citep{Mommert_2017}. Each filter was observed sequentially, with each sequence consisting of twelve $90\,$s exposures, equivalent to one full rotation of YR4, thereby averaging over the lightcurve amplitude. These observations complement the Sloan‐band data by extending visible coverage to slightly different central wavelengths. We used \citet{Holmberg_2006} colors of the Sun to calibrate our observations obtained in the Johnson–Cousins filters. 

Near-infrared observations were acquired with the HAWK-I instrument at the VLT using the Y, J, and H filters. The instrumental magnitudes were first calibrated using field stars from the 2MASS catalog \citep{Skrutskie_2006}. As the HAWK-I filter set differs slightly from the 2MASS system, we then applied the color transformations derived by \citet{Coccato_2021} to convert our calibrated photometry onto the 2MASS photometric system, and used the same transformations to convert the solar absolute magnitudes from \citet{Willmer_2018}. The data were obtained following a YHHYJ exposure sequence, with each segment lasting approximately 20 minutes, corresponding to one full rotation of the object, and consisting of multiple individual exposures. The images from each filter were co-added to enhance the signal-to-noise ratio, and photometry was performed on the resulting stacked images.

To calibrate the HAWK-I data, we also used quasi-simultaneous R-band observations ($R_{\rm SPECIAL}$ filter) from FORS2 (also at the VLT) as a visible anchor. These were used, and processed using a similar strategy, to normalize the NIR fluxes and ensure consistency across the combined spectrum. Although the LDT visible observations were obtained at a different epoch, the FORS2 R-band photometry enabled us to rescale the near-infrared fluxes and directly compare them with the DK154, LDT and TNG reflectance data.

\begin{figure*}[t!]
\centering
\includegraphics[width=13cm]{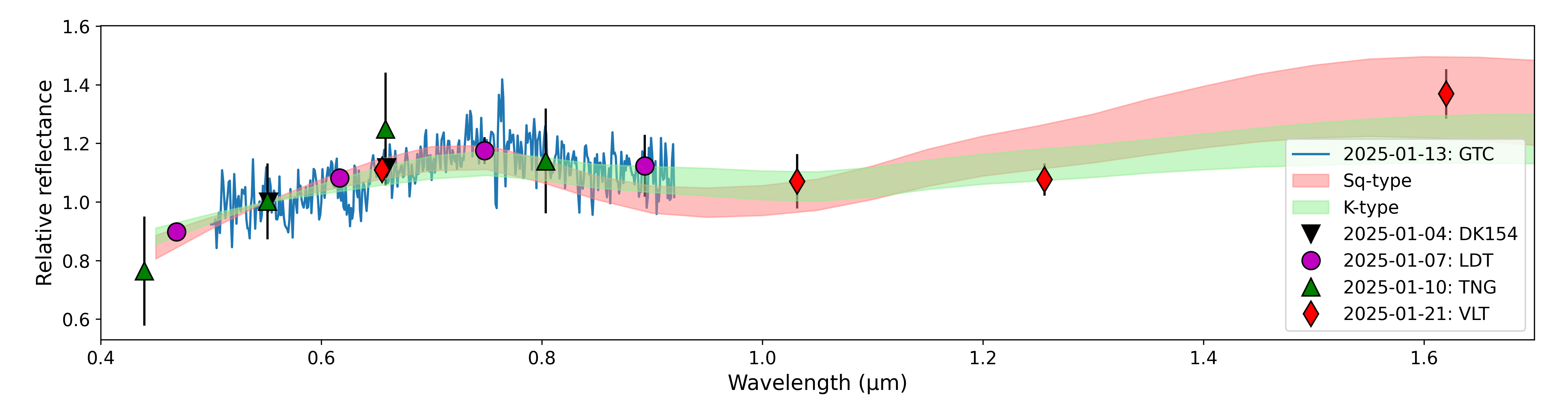}
\caption{Relative reflectance spectrum and colors of YR$_4$. The reflectance was normalized to 1 at 0.55$\mu$m}.
\label{fig:Colors_Spec}
\end{figure*}

The combined visible and near-infrared reflectance spectrum is presented in Fig.~\ref{fig:Colors_Spec}, normalized to unity at the V band. This spectrum was subsequently used for taxonomic classification (see Sect.~\ref{sec:taxonomy}). All the colors derived for the different telescopes are listed in Table \ref{tab:color_indices}.

\begin{table}[ht]
\centering
\caption{Color indices for YR4 from various telescopes}
\label{tab:color_indices}
\begin{tabular}{lccc}
\hline
Telescope & Color  & Color (mag) & Uncertainty (mag) \\
\hline
DK154   & $V-R$ &  0.48 & 0.02 \\
LDT     & $g-r$ & 0.20 & 0.05 \\
LDT     & $r-i$ & 0.09 & 0.05 \\
LDT     & $i-z$ & $-0.05$ & 0.11 \\
TNG     & $B-V$ &  0.94 & 0.26 \\
TNG     & $V-R$ &  0.60 & 0.17 \\
TNG     & $R-I$ & 0.24 & 0.13 \\
VLT     & $R-Y$ &  0.46 & 0.10 \\
VLT     & $Y-J$ &  0.30 & 0.12 \\
VLT     & $J-H$ &  0.58 & 0.09 \\
\hline
\end{tabular}
\end{table}

\subsection{GTC visible spectroscopy}
\label{ssec:vis_spec}
Low-resolution visible spectroscopy of YR4 was obtained with the OSIRIS+ instrument on the GTC. To obtain the reflectance spectrum of the asteroid, two solar analog stars from the Landolt catalog \citep{Landolt_1992}, SA98-978 and SA102-1081, were also observed at a similar airmass to that of the asteroid and using the same instrumental configuration. Data reduction was done using standard procedures. Images were first bias and flat-field corrected, and sky subtracted. Next, a one-dimensional spectrum was extracted from the two-dimensional images using an aperture corresponding to the pixel where the intensity decayed to 10\% of the peak intensity value. Wavelength calibration was applied to the 1D-spectra using Xe+Ne+HgAr lamps, and the same procedure was applied to the spectra of the stars. We divided the asteroid's individual spectra by the spectra of the solar analogs, and the resulting ratios were finally averaged to obtain the reflectance spectrum of YR4. As shown in Fig.~\ref{fig:Colors_Spec}, the OSIRIS spectrum aligns well with the reflectance values derived from broadband photometry. 

\subsection{Taxonomic classification}
\label{sec:taxonomy}
We analyzed the combined spectrum using the Bus-DeMeo taxonomy \citep{Demeo_2009}. The photometric spectrum was resampled to the mean spectral templates of each class and compared using a least-squares approach.

Using the Bus-DeMeo system, the best-fit match is the Sq-type, followed closely by S- and K-type solutions. However, recent JWST thermal-infrared observations suggest an equivalent diameter for YR4 of $60\pm7$\,m \citep{Rivkin_2025} resulting in an albedo of $\sim 0.11$ for YR4 if we take into account the $H=24.14$ mag derived in this work. Such an albedo value is somewhat lower than typical values for Sq-type asteroids, but corresponds to what we would be expecting for a K-type object. 

\section{Discussion}

\subsection{A real-time planetary defense scenario}

The case of asteroid YR4 represents a genuine planetary defense scenario, not a simulation as those conducted during the Planetary Defense Conferences (PDC) \citep{Chodas_2019} or the IAWN exercise observation campaigns \citep{Reddy_2019,Reddy_2022a,Reddy_2022b,Reddy_2024,Farnocchia_2022,Farnocchia_2023}. Following its discovery on 27 December 2024, YR4 was identified as a virtual impactor with a non-negligible impact probability for December 2032. Over the following weeks, as new observations refined the orbit, the impact probability steadily increased. On 27 January 2025, it reached Torino Scale 3, the first time an asteroid had ever been formally classified at this level. This milestone also triggered the first-ever public notification from the International Asteroid Warning Network (IAWN), reflecting the seriousness of the risk.

\subsection{Timeline and coordination of the observational response}

Observations were initiated and executed under real-time threat conditions, not for demonstration purposes but as a necessary response to the evolving impact risk. The timeline of observational efforts presented in this paper closely followed the evolution of the impact probability. We only present observations obtained for physical characterization and intentionally left out observations from the James Webb Space Telescope that will be presented in other papers. The timeline is the following:

\begin{itemize}
    \item 28 December (Cumulative impact probability: 1/1621): Catalina Sky Survey (CSS) obtained high-cadence imaging one day after discovery under their regular survey strategy of the sky.
    \item 3-4 January (Cumulative impact probability: 1/1041): The Danish 1.54\,m telescope obtained two light curves and determined for the first time the rotation period. 
    \item 7 January (Cumulative impact probability: 1/961): As the impact probability was increasing, request was made for observations at the 4.3\,m LDT. Broad-band colors and light-curve data were consequently collected via the MANOS program.
    \item 10 January (Cumulative impact probability: 1/974): A DDT request under the NEOPOPS project was executed at the TNG, providing visible-color photometry.
    \item 13 January (Cumulative impact probability: 1/965): Visible spectroscopy was acquired at the GTC under an ongoing regular program focused on high-priority NEOs.
    \item 20-21 January (Cumulative impact probability: 1/327): After another significant increase of the impact probability, near-infrared observations with HAWK-I at the VLT were requested and obtained via a DDT, while simultaneous R-band photometry with FORS2 was secured through the ESA–ESO planetary defense partnership.
    \item 30 January to 2 February  (Cumulative impact probability: 1/78): Observations at the NOT were obtained and used to improve the phase curve coverage just as the object approached peak concern.
    \item 18 February: YR4 cumulative impact probability reached its highest value at 2.8\% according to the ESA Aegis system, 3.1\% according to JPL Sentry, and 2.3\% according to the CLOMON2 system. However, at that time, YR4 had become too faint to obtain new meaningful physical characterization observations from ground based facilities.
    
\end{itemize}

\subsection{Rapid physical characterization under uncertainty}

The observations performed in this work were obtained while YR4 was still classified as a potential impactor, and most were completed before or during the period in which the object was at Torino Scale 3 (27 January to 18 February). The campaign was not a validation exercise, it was carried out under urgent, operational conditions.

The physical data collected, rotation period, taxonomy, size \citep[from][]{Rivkin_2025}, albedo \citep[based on the size from][]{Rivkin_2025}, and phase behavior, were essential for supporting risk assessments and informing mitigation readiness. 

With a rotation period of $P = 19.46341 \pm 0.00008$\,min YR4 is considered as a fast rotator. However, the vast majority of small asteroids are fast rotators as this adjective is used to contrast with objects that rotate with period around or larger than two hours, known as the spin barrier \citep{Pravec_2000}. Small NEOs ($D<150$\,m) can rotate faster than the spin barrier as small internal cohesion allow them to maintain their structure even under the stress experienced by their fast rotation \citep{Sanchez_2014}.

\begin{figure}[t!]
\centering
\includegraphics[width=8cm]{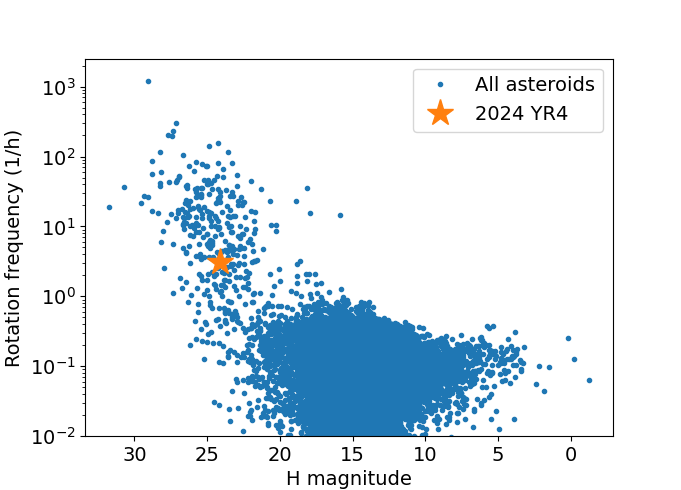}
\caption{Rotation frequency ($1/P$) versus absolute magnitude $H$ for asteroids with published spin periods retrieved from the \textit{rocks} database (blue dots; only the first reported period per object is used). The orange star marks the location of YR4.}
\label{fig:spin_H}
\end{figure}

Figure~\ref{fig:spin_H} places YR4 within the broader small-body population by plotting rotation frequency versus absolute magnitude for all asteroids in the \textit{SsODNet} database \citep{Berthier_2023} that have at least one published period. Objects larger than \(\sim\!150\) m (right side; $H<22$) cluster below the 2.2\,h spin barrier, whereas sub-150\,m bodies ($H>22$) form a cloud of fast rotators extending to hundreds of revolutions per day. The orange star shows that YR4 sits near the center of this cloud, indicating that its $19.46341 \pm 0.00008$\,h period is typical for an asteroid of its size. Such a rotation period over the spin barrier limit implies that YR4 needs some internal strength to hold itself against the acceleration induced by rotation, which overcomes the gravitational one. However, this does not imply that YR4 needs to be monolithic as the needed internal strength is very low ($\sim$ few tenth of Pa) \citep{Sanchez_2014,Holsapple_2007}. 

Highly accurate knowledge of the rotation period and shape of the object is extremely important for planetary defense. A deflection or a fly-by reconnaissance mission might need to target specific rotation phases. In the case of an impact on the asteroid for a deflection or a disruption of the object, while the rotation period has little effect on the deflection result, it has a major impact on the probability to disrupt the object or not \citep{Syal_2016}. The faster the asteroid rotates, the higher the probability of disruption is \citep{Syal_2016}. A large number of lightcurve observations also provide information on the shape of the object, which can also be a critical parameter for any deflection/disruption mission \citep{Syal_2016}. Moreover, a rendezvous mission might want to orbit the object and thus need a precise shape model.

Despite the urgency of the threat assessment, the campaign succeeded in securing the smallest phase angle that was observationally accessible after discovery: $\alpha = 7.4^{\circ}$, obtained with FORS2 only three weeks after the first detection. Although this value is close to the practical limit imposed by YR4's 2025 observing geometry, it remains too large to constrain the opposition–surge regime. Consequently, the $H$ magnitude is poorly constrained and the size estimate derived from $H$ and an assumed albedo carries large uncertainty. This limitation is non-trivial for planetary defense work, where $H$ often provides the only rapid proxy for an object's diameter and impact energy.  

The next opportunities to reach lower phase angles occur during the 2028–2029 apparition.  Prior to closest approach, YR4 will reach a minimum of $\alpha = 1.9^{\circ}$, but at a faint $V \simeq 24.9$\,mag. Post-fly-by geometry is less favorable but the object will be brighter. On 29 January 2029 the asteroid attains $\alpha = 6.2^{\circ}$ at $V \simeq 22.4$\,mag.

Subsequent {\em JWST} MIRI observations obtained in March~2025 yielded a thermal–infrared diameter of $D_{\rm eq}=60\pm7$\,m \citep{Rivkin_2025}. Combining this size with the range of absolute magnitudes derived above ($H=24.14\pm0.25$\,mag) implies a geometric albedo of $p_{\rm V} = 0.11 \pm 0.05$.

Such an albedo is on the low side for ordinary S-complex objects but would be fully consistent with a K-type surface. In contrast, the shallow phase slope ($G=0.51$) would normally be interpreted as evidence for a high-albedo regolith. \citep{Shevchenko_2019} found a correlation between the phase integral $q$ and the albedo of asteroids. According to \citet{Bowell1989}, $q$ can be derived from $G$ from the relation $q = 0.290 + 0.684G$. Substituting our $G=0.51$, we find $q=0.639$ which has been found to be related to high albedo object ($P_{\rm V} = 0.45 \pm 0.07$) according to \citep{Shevchenko_2019}. The mismatch between the thermally inferred size, the $H$-derived albedo, and the phase‐curve slope leaves YR4's taxonomic classification and $H$ magnitude ambiguous.

Polarimetric observations would have offered an independent albedo estimate. The degree of linear polarization measured at large phase angles ($\alpha \gtrsim 40^{\circ}$) is inversely correlated with the geometric albedo. Such data would have provided an independent estimation of the diameter (combined with the $H$ magnitude) from the one derived by the JWST. Unfortunately, during the 2024–2025 apparition YR4 was only observable at such a high phase angle shortly after discovery.

The determination of the albedo would have been of significant importance. The visible-spectrum shape favors an Sq classification, while the moderate albedo and ambiguity in the NIR spectrum allow for a K-type interpretation. Based on its spectral characteristics, YR4 most plausibly falls into either the Sq- or K-taxonomic classes. This distinction is important for planetary defense analyses because the two classes have different meteorite analogues and therefore different typical bulk densities. Sq/S-type objects are linked to ordinary chondrites (H, L, LL) \citep{Dunn_2010,Nakamura_2011}, whose laboratory bulk densities cluster around $3.3$–$3.4~\mathrm{g\,cm^{-3}}$ \citep{Britt_Consolmagno_2003}. K-type asteroids, by contrast, correspond to CO/CV (and CK) carbonaceous chondrites \citep{Bell_1988,Burbine_2001}, which are appreciably lighter at roughly $2.9$–$3.0~\mathrm{g\,cm^{-3}}$ \citep{Macke_Consolmagno_Britt_2011}. The determination of the mass of YR4 is out of the scope of this paper as asteroid densities can vary significantly from meteorite bulk densities as their macroporosity can be significant and has not been estimated for YR4. However, all else being equal (i.e., assuming comparable size and macroporosity), a K-type composition would imply a $\sim10\%$ lower mass and hence a less energetic impact—than an Sq/S-type analogue.

These characteristics would have directly informed any modeling of impact consequences and response strategies. Importantly, this information was made available while the asteroid remained an open impact risk, illustrating the value of having coordinated observing resources and analysis pipelines ready to act on short notice.

\subsection{Implications for planetary defense readiness}

The YR4 campaign highlights the critical importance of having both infrastructure and operational procedures in place before a crisis emerges. Telescope access mechanisms, such as standing ESA–ESO partnerships and fast DDT approval paths, were crucial in securing rapid follow-up. Impact monitoring systems (Sentry, Aegis, CLOMON2) provided consistent tracking of the evolving risk. Communication channels, including IAWN, functioned as intended.

Although the terrestrial impact risk was ultimately ruled out, a $\sim4\%$ probability of collision with the Moon in 2032 remains and is higher than the peak Earth impact probability. Thus, YR4 continues to be a high-priority target for follow-up in the coming years. Recent works even suggested that a lunar impact could significantly increase the rate of micro-meteorite impacts to artificial satellites around Earth. Thus, lunar impacts should also be considered in the planetary defense umbrella \citep{Wiegert_2025}. The rapid characterization achieved here, on a timescale of days to weeks, should serve as a benchmark for future planetary defense responses, demonstrating what can be done when observations are driven by true operational necessity rather than by a predefined test scenario.

\section{Conclusions}

We presented an intensive observational campaign focused on asteroid 2024~YR$_4$, initially classified as a potential Earth impactor and the first object to reach Torino Scale 3. Rapid-response photometric and spectroscopic observations from multiple facilities and instruments (LDT, TNG, VLT, NOT, GTC, and archival CSS data) allowed us to characterized its physical properties, providing crucial information for planetary defense assessments.

Our main results can be summarized as follows:

\begin{enumerate}
    \item We determined a precise synodic rotation period of $19.46341 \pm 0.00008$\,min, placing 2024~YR$_4$ among rapidly rotating near-Earth asteroids. Such fast rotation is consistent with known small-body populations and suggests typical cohesive strength characteristics for an object of its size.
    
    \item The combined photometric and spectroscopic analysis indicates that 2024~YR$_4$ is most consistent with an Sq- or K-type taxonomic classification.
    
    \item The phase curve analysis yielded a shallow slope parameter ($G = 0.51\pm 0.11$) and an absolute magnitude of $H_\mathrm{R} = 23.82$. After color correction and considering other models for the phase curve analysis we obtain a final estimate of the $H$ magnitude of $24.14 \pm 0.25$. This relatively high $G$ value is suggestive of a high albedo surface while the $H$ magnitude combined with the JWST observations are suggestive of a low albedo object. A better constraint of the $H$ magnitude would be needed to understand this discrepancy which requires observations closer to opposition that can only be obtained shortly before and after the next fly-by in 2028. Another solution to this discrepancy would be to obtain an independent estimation of the 2024~YR$_4$ albedo through polarimetric observations.
    
    \item Crucially, our rapid observational response and timely characterization provided essential inputs to refine the physical properties of YR4. The successful international coordination and prompt data acquisition demonstrated during this campaign validate current planetary defense procedures and underscore the importance of preparedness and rapid-response capabilities.
\end{enumerate}

Although 2024~YR$_4$ was eventually downgraded from its initial impact-risk with Earth, this event provided a valuable opportunity to test and refine planetary defense observational and analytical strategies. Moreover, 2024~YR$_4$ remains a high-priority target for planetary defense, as it still has an $\sim4\%$ probability of impacting the Moon in 2032. Such an impact could create a large amount of debris in Earth orbit and could potentially be harmful for human assets in orbit around Earth. Continued investment in rapid-response observational networks and characterization capabilities will be critical in effectively managing potential asteroid impact threats in the future.

\backmatter

\bmhead{Acknowledgments}

Based on observations collected at the European Southern Observatory under 
ESO programmes 113.2690.002 and 114.28HT.001.
The authors are grateful to ESO for the extremely fast response time for our DDT proposal.
The observations at the 1.54m Danish telescope at La Silla Observatory in Chile and the work at Ond\v{r}ejov were supported by the {\it Praemium Academiae} award (no. AP2401) from the Academy of Sciences of the Czech Republic.
Based on observations made with the Nordic Optical Telescope, owned in collaboration by the University of Turku and Aarhus University, and operated jointly by Aarhus University, the University of Turku and the University of Oslo, representing Denmark, Finland and Norway, the University of Iceland and Stockholm University at the Observatorio del Roque de los Muchachos, La Palma, Spain, of the Instituto de Astrofisica de Canarias.
The NOT data were obtained under program ID P68-803).
Based on observations collected at the Telescopio Nazionale Galileo (TNG, A50TAC31), operated on the island of La Palma by the Centro Galileo Galilei of the INAF (Istituto Nazionale di Astrofisica) at the Spanish Observatorio del Roque de los Muchachos of the Instituto de Astrofísica de Canarias.
The 4.3m Lowell Discovery Telescope observations have been obtained thanks to the support of the NASA YORPD grant 80NSSC21K1328 awarded to the Mission Accessible Near-Earth Object Survey (MANOS).
The NEOPOPS observations were funded by a programme of the European Union and implemented by ESA (agreement No.4000147191/25/D/MRP).

\begin{itemize}

\item Author contribution: M.D. wrote the main manuscript text, prepared all the figures, reduced and analyzed the, FORS2, NOT, and CSS observations.
M.D., O.H., and M.M. wrote the proposal to obtain the VLT HAWK-I observations. 
M.D. and O.H. reduced and analyzed the HAWK-I observations.
P.P. obtained, reduced, and analyzed the observations from the Danish 1.54m telescope. 
J.L.C., L.C., and F.O. substantially participated to the preparation of the manuscript.
N.M. reduced and analyzed the 4.3m LDT observations.
J.de L. obtained, reduced and analyzed the GranTeCan observations.
Z.G., G.F. and M.K. wrote the proposal to the NOT and reduced the observations.
J.B. reduced and analyzed the TNG observations.
S.I. wrote the proposal to obtain the TNG observations.
E.D. is the PI of the NEOPOPS consortium that obtained the TNG observations.
T.B. and C.F. performed the observations at the Catalina Sky Survey
T.K. performed the observation at the Lowell Discovery Telescope
P.P, J.B., S.I, E.D., S.B., M.A.B., M.B., A.F., K.H., P.H., P.F., P.K., F.F., M.F., M.L., F.L.F., E.M.E., A.M., D.P., P.B., J.P.R., and A.C  are part of the NEOPOPS consortium 
All. authors reviewed the manuscript.
\end{itemize}

\bibliography{sn-bibliography}

\end{document}